\documentclass[preprint]{aastex}

\begin{document}

\title {KIC 4739791: A New R CMa-type Eclipsing Binary with a Pulsating Component }
\author{Jae Woo Lee$^{1,2}$, Seung-Lee Kim$^{1,2}$, Kyeongsoo Hong$^1$, Jae-Rim Koo$^1$, Chung-Uk Lee$^{1,2}$, and Jae-Hyuck Youn$^1$ }
\affil{$^1$Korea Astronomy and Space Science Institute, Daejeon 305-348, Korea}
\affil{$^2$Astronomy and Space Science Major, Korea University of Science and Technology, Daejeon 305-350, Korea}
\email{jwlee@kasi.re.kr, slkim@kasi.re.kr, kshong@kasi.re.kr, koojr@kasi.re.kr, leecu@kasi.re.kr, jhyoon@kasi.re.kr}

\begin{abstract}
The {\it Kepler} light curve of KIC 4739791 exhibits partial eclipses, inverse O'Connell effect, and multiperiodic pulsations. 
Including a starspot on either of the binary components, the light-curve synthesis indicates that KIC 4739791 is in detached 
or semi-detached configurations with both a short orbital period and a low mass ratio. Multiple frequency analyses were performed 
in the light residuals after subtracting the binarity effects from the original {\it Kepler} data. We detected 14 frequencies: 
six in the low-frequency region (0.1$-$2.3 d$^{-1}$) and eight in the high-frequency region (18.2$-$22.0 d$^{-1}$). Among these, 
six high frequencies with amplitudes of 0.62$-$1.97 mmag were almost constant over time for 200 d. Their pulsation periods and 
pulsation constants are in the ranges of 0.048$-$0.054 d and 0.025$-$0.031 d, respectively. In contrast, the other frequencies 
may arise from the alias effects caused by the orbital frequency or combination frequencies. We propose that KIC 4739791 is 
a short-period R CMa binary with the lowest mass ratio in the known classical Algols and that its primary component is 
a $\delta$ Sct pulsating star. Only four R CMa stars have been identified, three of which exhibit $\delta$ Sct-type oscillations. 
These findings make KIC 4739791 an attractive target for studies of stellar interior structure and evolution.
\end{abstract}

{}

\section{INTRODUCTION}

Asteroseismology is used to investigate the interior structure of stars through their pulsation features. Among them, $\delta$ Sct
and $\gamma$ Dor pulsators with spectral types A-F are of particular interest to us. They have observational properties that are 
very similar to each other but that exhibit significant differences in pulsation periods and pulsation constants 
(Handler \& Shobbrook 2002). $\delta$ Sct stars pulsate in low-order press ($p$) modes driven by the $\kappa$ mechanism with 
short periods of 0.02$-$0.2 d and small pulsation constants of $Q <$ 0.04 d (Breger 2000), whereas $\gamma$ Dor stars pulsate in 
high-order gravity ($g$) modes caused by the convective blocking mechanism with relatively longer periods of 0.4$-$3 d and 
larger pulsation constants of $Q >$ 0.23 d (Kaye et al. 1999, Henry et al. 2005). Although $\delta$ Sct pulsators are generally 
hotter than those of $\gamma$ Dor, the overlap in their instability strips indicates the possible existence of hybrid stars 
exhibiting both $\gamma$ Dor and $\delta$ Sct pulsations. Such pulsators are of significant interest because the $g$ modes assist 
in probing the deep interiors of a star and the $p$ modes in probing its envelope. A recent study from {\it Kepler} photometry 
indicates that nearly all $\delta$ Sct stars pulsate in low frequencies of 0$-$5 d$^{-1}$ and that they are 
$\delta$ Sct/$\gamma$ Dor hybrids (Balona 2014). 

The pulsating components in eclipsing binaries (EBs) are attractive objects that exhibit eclipses and pulsations simultaneously. 
They allow accurate and direct determination of fundamental parameters such as mass and radius for each component, and they 
provide an important constraint on the stellar structure and evolution models. Furthermore, the effects of tidal forces on 
pulsations can be investigated. Many of them have been found to be $\delta$ Sct-type members of the classical semi-detached Algols 
designated as oscillating EA (oEA) objects; Mkrtichian et al. (2004) identified the new class of $\delta$ Sct stars as the (B)A-F 
spectral type mass-gaining components of Algol systems. The oEA stars have nearly the same pulsating characteristics as 
$\delta$ Sct stars, but they have experienced a very different evolution process by tidal interaction and mass transfer between 
the binary components. Recently, numerous candidate hybrid stars that exhibit two types of pulsations have been detected from 
space missions, but only four stars have been identified as EBs containing the hybrid $\gamma$ Dor/$\delta$ Sct components: 
CoRot 100866999 (Chapellier \& Mathias 2013), KIC 4544587 (Hambleton et al. 2013), KIC 3858884 (Maceroni et al. 2014), and 
KIC 8569819 (Kurtz et al. 2015). Their physical parameters can be well defined from detailed studies of the EBs, which makes them 
excellent targets for asteroseismology and the study of stellar interior and evolution.

We have been looking for pulsating components in EBs using the highly precise {\it Kepler} data. In our first result, we discovered 
a quadruple star system V404 Lyr (KIC 3228863) that exhibits $\gamma$ Dor pulsations (Lee et al. 2014). In this paper, we analyze 
in detail the {\it Kepler} photometric observations of KIC 4739791 
(RA$_{2000}$=19$^{\rm h}$18$^{\rm m}$50$\fs5$; DEC$_{2000}$=+39$^{\circ}$49${\rm '}$04$\farcs$8; $K_{\rm p}$=$+$14.703) 
with about 0.8989 d binary period and present that it is an R CMa-type eclipsing system with a pulsating component.

\section{{\it KEPLER} PHOTOMETRY AND LIGHT-CURVE SYNTHESIS}

The {\it Kepler} photometry of KIC 4739791 has been performed during Quarters 14 and 15 in the long cadence mode with 29.42-min 
integration times. We used the data in the {\it Kepler} EB catalogue\footnote{http://keplerebs.villanova.edu/}, which were 
detrended and normalized from the raw SAP (Simple Aperture Photometry) time series. Detailed descriptions of the catalogue 
can be found in Pr\v sa et al. (2011) and Slawson et al. (2011). The light curve of the binary system is plotted in Figure 1 
as the normalized flux versus the orbital phase. Its shape is the familiar one for classical Algol-type binaries and therefore 
signifies a significant temperature difference between the two components. Furthermore, the {\it Kepler} data are asymmetrical 
and have Max I fainter than Max II by amounts of approximately 0.004 mag. This might result from local photospheric 
inhomogeneities and could be explained by starspot activity on the components. 

In order to derive the physical parameters of KIC 4739791, we set the mean light level at phase 0.25 to unity and simultaneously 
analyzed all individual observations using the 2007 version of the Wilson-Devinney binary code 
(Wilson \& Devinney 1971, van Hamme \& Wilson 2007; hereafter W-D). The light-curve synthesis was performed in a manner almost 
identical to that for the eclipsing systems V404 Lyr (Lee et al. 2014) and KIC 5621294 (Lee et al. 2015). The effective temperature 
of the brighter and presumably more massive star was initialized at $T_{1}$=7,761 K from the revised stellar properties of 
the {\it Kepler} targets recently compiled by Huber et al. (2014). In Table 1, the primary and secondary stars refer to those 
being eclipsed at Min I and Min II, respectively, and the parentheses indicate the adjusted parameters: the orbital ephemeris 
($T_0$ and $P$), the mass ratio ($q$=$M_2/M_1$), the inclination angle ($i$), the temperatures ($T_{1,2}$) and 
dimensionless surface potential ($\Omega_{1,2}$) of both components, and the monochromatic luminosity ($L_{1}$).

The mass ratio $q$ is a very important parameter because the W-D code is based on the Roche geometry which is sensitive to it. 
However, because there is no spectroscopic measurements, we conducted a photometric $q$-search procedure for various modes of 
the W-D code (Wilson \& Biermann 1976) to understand the geometrical structure and the photometric parameters of the system. 
The results only converged satisfactorily when mode 2 (detached system) and mode 5 (semi-detached system with the secondary 
filling its inner Roche lobe) were used; the other modes failed to achieve convergence. As depicted in Figure 2, 
the minimum values of $\Sigma$ (the weighted sum of the residuals squared, $\Sigma W(O-C)^2$) occurred around $q$=0.20 and 
$q$=0.07 for the detached and semi-detached modes, respectively. Then, the initial values of $q$ were treated as 
an adjustable parameter in the subsequent calculations in order to obtain the photometric solutions. The unspotted solution 
for the semi-detached mode is plotted as a dashed curve in the top panel of Figure 1. The light residuals are plotted in 
the middle panel of the figure, wherein it can be seen that this model does not describe the observed light curve satisfactorily. 
The result for the detached model also led to the same results.

The light discrepancy might be attributed to either magnetic cool spots on the secondary star with a deep convective envelope 
or a hot spot on the surface of the primary star as a result of the impact of a gas stream from the companion. However, there is, 
at present, no way to know which mechanism is more appropriate to explain the light change. Thus, we tested three possible spot 
models: a cool spot on the secondary star in both modes and a hot spot on the primary star in the semi-detached mode. 
The final results are given in Table 1 together with the spot parameters. Although it is difficult to distinguish between 
the spot models, as well as the Roche configurations, from only the light-curve analyses, the hot-spot model in 
a semi-detached state describes the {\it Kepler} data quite well and provides a smaller value of $\Sigma W(O-C)^2$ than 
the cool-spot models in both modes. The synthetic curve from the hot-spot model is plotted as a solid curve in Figure 1, and 
the residuals from the spot solution are plotted in the bottom panel of the figure. In all procedures before and after 
the spot parameters were evaluated, we included the orbital eccentricity ($e$), the argument of periastron ($\omega$), and 
the third light\footnote{This parameter is used to separate the light contribution of a third star from the overall brightness 
budget. The third-light source may be nearby stars which are only optically related with binary systems or circumbinary companions 
physically bound to them.} ($\ell_3$) as additional free parameters. The searches of $e$ and $\ell_3$ led to values for 
the two parameters which remained zero within their margins of errors. This indicates that KIC 4739791 is in a circular orbit.

From its temperature, the primary star of KIC 4739791 was assumed to be a normal main-sequence one with a spectral type of 
about A7. Because the temperature errors given in Table 1 are probably underestimated, it was assumed that the temperature of 
each component had an error of 200 K. Using the correlations between spectral type and stellar mass, we estimated 
the primary's mass to be $M_1$=1.75$\pm$0.08$M_\odot$. The absolute dimensions of the system can be calculated from 
our photometric solutions and $M_1$. These are given in the entries in the last part of Table 1, where the primary star is 
the hotter, larger, and more massive component. The luminosities ($L$) and bolometric magnitudes ($M_{\rm bol}$) were computed 
through adopting $T_{\rm eff}$$_\odot$=5,780 K and $M_{\rm bol}$$_\odot$=+4.73 for the solar values.

\section{LIGHT RESIDUALS AND PULSATIONAL CHARACTERISTICS}

In Figure 3, the light curve residuals from our hot-spot model versus BJD are plotted, wherein the lower panel presents 
a short section of the residuals. Light variations with a semi-amplitude larger than 5 mmag are clearly seen in the residuals 
and their amplitudes appear to vary with time. From the temperature ($T$) and surface gravity ($\log$ $g$) given in Table 1, 
the primary star of KIC 4739791 lies within the $\delta$ Sct instability strip, which is slightly hotter than the blue edge of 
the $\gamma$ Dor region; hence, it is a candidate for $\delta$ Sct and/or $\gamma$ Dor pulsators. Gaulme \& Guzik (2014) 
suggested that the system is pulsating at $\delta$ Sct frequencies of 200$-$283 $\mu$HZ, which corresponds to approximately 
17$-$24 d$^{-1}$. Using the PERIOD04 program (Lenz \& Breger 2005), we applied multiple frequency analyses to only 
the outside-eclipse residuals (orbital phases 0.08$-$0.42 and 0.58$-$0.92) after eliminating the data of both eclipses in order 
to obtain more reliable frequency analyses. The top panel of Figure 4 displays the amplitude spectra in the frequency range 
from 0 to the Nyquist limit of $f_{\rm Ny}$=24.47 d$^{-1}$. We can see that the main signals lie in two frequency regions: 
$<$6.0 d$^{-1}$ and $>$16.0 d$^{-1}$.

As a result of the successive pre-whitening procedures, we detected a total of 14 frequencies with signal-to-noise amplitude 
(S/N) ratios larger than 4.0 (Breger et al. 1993). At each procedure, we applied a multiperiodic least-squares fit 
to the light residuals using the equation of $Z$ = $Z_0$ + $\Sigma _{i}$ $A_i \sin$(2$\pi f_i t + \phi _i$). Here, $Z$ and 
$Z_0$ denote the calculated magnitude and zero point, respectively; $A_i$ and $\phi _i$ are the amplitude and phase of 
the $i$th frequency, respectively; and $t$ is the time of each measurement. The amplitude spectra after pre-whitening 
the first seven frequencies and then all 14 frequencies are presented in the second and third panels of Figure 4, respectively. 
The synthetic curve computed from the 14-frequency fit is displayed in Figure 3, and the results are listed in Table 2. 
The uncertainties in the table were derived according to Kallinger et al. (2008). Of the 14 frequencies, six are in 
the lower frequency $g$-mode region and eight are in the $p$-mode region. 

The {\it Kepler} observations were obtained in the long-cadence (LC) mode of 29.42-min and/or the short-cadence (SC) mode of 
58.85 s (Koch et al. 2010). For each mode, 6.02-s exposures with their associated 0.52-s readout times were co-added on board: 
270 times for LC and nine times for SC. Thus, the frequencies detected in the LC data could be partly affected by 
the merging effect. In order to examine whether the pulsation frequencies of KIC 4739791 were real or aliased, we created 
artificial data calculated with 6.54-s sampling rates from the 14 frequencies in Table 2 and then binned sets of 270 
consecutive points from those data to form the 29.42-min observations, just as in the case of the {\it Kepler} LC data. 
The periodogram for the simulated data is presented in the bottom panel of Figure 4. We can see that the frequencies were 
almost identical to those from the original {\it Kepler} LC data, while the amplitudes were reduced to less than 5 per cent in 
the low frequencies and 22$-$30 per cent in the high frequencies. On the other hand, the frequencies (18$-$22 d$^{-1}$) near 
the Nyquist frequency could be reflections of real frequencies (2$f_{\rm Ny}-f_i$) higher than $f_{\rm Ny}$ (Murphy et al. 2013). 
Precise photometry with high time resolutions would identify the pulsation frequencies detected in this paper.

We examined the frequency variations with time through analyzing the outside-eclipse residuals at intervals of $\sim$50 d. 
The four subsets resulted in slight differences from each other, and an average of 10 frequencies were detected at each subset. 
Figure 5 displays the variability of the eight frequencies detected most repeatedly. The six frequencies of $f_2$, 
$f_5-f_7$, and $f_9-f_{10}$ were almost constant with time, while the $f_1$ and $f_{8}$ frequencies varied significantly. 
The variations of the two low frequencies might be partially attributed to changes in the spot parameters over time caused 
by local photospheric inhomogeneities. Within the frequency resolution of 0.008 d$^{-1}$ (Loumos \& Deeming 1978), $f_1$ and 
$f_8$ appeared to be the orbital frequency ($f_{\rm orb}$=1.11245 d$^{-1}$) and its multiple, respectively, and $f_3$ and 
$f_4$ were the equally spaced sidelobes split from $f_1$ by 0.00533 d$^{-1}$. The lowest frequency signal $f_{13}$ might
result from either a combination frequency of $f_2-f_7-f_1$ or a long-term trend that resides in the data.

\section{DISCUSSION AND CONCLUSIONS}

In this paper, we presented the physical properties of KIC 4739791 obtained from detailed analyses of the {\it Kepler} data.
The light curve with Max II brighter than Max I was satisfactorily modelled through applying a single spot to either of 
the components. This synthesis indicates that the binary system is in detached or semi-detached states with 
a short orbital period and a very low mass ratio. In the detached mode, the primary and secondary components fill $F_1$=74 \% 
and $F_2$=79 \% of their limiting lobe, respectively, while $F_1$=64 \% in the semi-detached mode. Here, 
$F_{1,2}$=$\Omega_{1,2}$/$\Omega_{\rm in}$, where $\Omega_{\rm in}$ is the potential of the inner critical surface in 
Roche geometry. These properties closely resemble those of the R CMa-type stars (Budding \& Butland 2011; Lehmann et al. 2013), 
which are characterized by relatively low $q$ and $P$ combinations among the Algol binaries. Two (R CMa and AS Eri) of 
the three R CMa stars exhibited $\delta$ Sct-type oscillations similar to oEA stars.

After removing the binarity effects from the original {\it Kepler} data, we performed multiple frequency analyses in 
the light residuals, using only the data in the out-of-eclipse phases. Fourteen pulsation frequencies with S/N ratios larger 
than 4.0 were found in two regions: 0.1$-$2.3 d$^{-1}$ and 18.2$-$22.0 d$^{-1}$. Among these, six frequencies ($f_2$, 
$f_5-f_7$, $f_9-f_{10}$) in the high-frequency region were highly stable for about 200 d, while two frequencies ($f_1$, $f_{8}$) 
in the low-frequency region varied significantly. We consider that most low frequencies are not the $g$-mode pulsations,
which might result from starspot activity or from alias effects caused by the orbital frequency. 
The $Q_D$ and $Q_{SD}$ for the high frequencies given in columns (6) and (7) of Table 2 are the pulsation constants for 
the detached and semi-detached solutions, respectively, which are well-known to be $Q$ = $P_{\rm pul}$$\sqrt{\rho / \rho_\odot}$, 
where $P_{\rm pul}$ is the pulsation period and $\rho$ is the mean density. The ratios of the pulsational-to-orbital periods 
for the $p$-mode frequencies were calculated to be $P_{\rm pul}/P_{\rm orb}$=0.05$\sim$0.06, which is within the upper limit 
of 0.09 for $\delta$ Sct stars in eclipsing binaries proposed by Zhang et al. (2013). The results demonstrate that 
the primary component of KIC 4739791 is a candidate for a $\delta$ Sct type pulsating star, rather than 
a hybrid $\delta$ Sct/$\gamma$ Dor pulsator. 

It is known that stars in close binaries evolve differently to single stars due to tidal interaction and probable mass 
transfer from their companions. From the absolute parameters of KIC 4739791, it is possible to estimate its evolutionary status 
in terms of the mass-radius, mass-luminosity, and mass-temperature diagrams. The locations of the components in these diagrams 
are presented in Figure 6, together with those of 137 well-studied Algol-type binaries, including the R CMa-type EBs 
(R CMa, AS Eri, OGLEGC 228). The data are taken from the compilations of Ibano\v{g}lu et al. (2006; 134 Algols), 
Kaluzny et al. (2007; OGLEGC 228), Budding \& Butland (2011, R CMa), and Lehmann et al. (2013; 
KIC 10661783\footnote{The eclipsing system is a detached binary with characteristics of the R CMa-type stars.}). 
In these diagrams, the primary component of KIC 4739791 is an almost unevolved star located on the zero-age main sequence (ZAMS), 
while the secondary star is highly evolved in a location where the secondary components of other Algol-type binaries barely exist. 
The secondary component is oversized by a factor of approximately two or six, and it is more than 20 or 56 times too luminous in 
the detached or semi-detached configurations, respectively, compared with its ZAMS mass. 

From the modelling of the {\it Kepler} light curve, we propose two possible solutions: a detached EB with a mass ratio of 
about 0.2 and a semi-detached EB with a mass ratio of about 0.07. If it is a semi-detached binary preferred in the light-curve 
analysis, KIC 4739791 is an Algol system with the lowest mass ratio ever observed and is considered an R CMa-type star that 
exhibits a very short orbital period. The secondary to primary mass transfer could be responsible for the hot spot on 
the mass-gaining primary component. Furthermore, the semi-detached nature would make KIC 4739791 a member of the class of 
oEA stars (Mkrtichian et al. 2004). Future high-resolution spectroscopy and multiband photometry will assist in establishing 
whether the system is a semi-detached EB and in revealing more accurate properties such as the absolute dimensions, 
identification of pulsation modes, and evolutionary status. Because KIC 4739791 is a faint star with a relatively short period,
the 8$-$10 m class telescopes such as VLT and Keck are needed to measure its accurate radial velicities.

\acknowledgments{ }
This paper includes data collected by the {\it Kepler} mission. {\it Kepler} was selected as the 10th mission of the Discovery Program. 
Funding for the {\it Kepler} mission is provided by the NASA Science Mission directorate. We have used the Simbad database maintained 
at CDS, Strasbourg, France. This work was supported by the KASI (Korea Astronomy and Space Science Institute) grant 2015-1-850-04.

\clearpage
\begin{figure}
\includegraphics[]{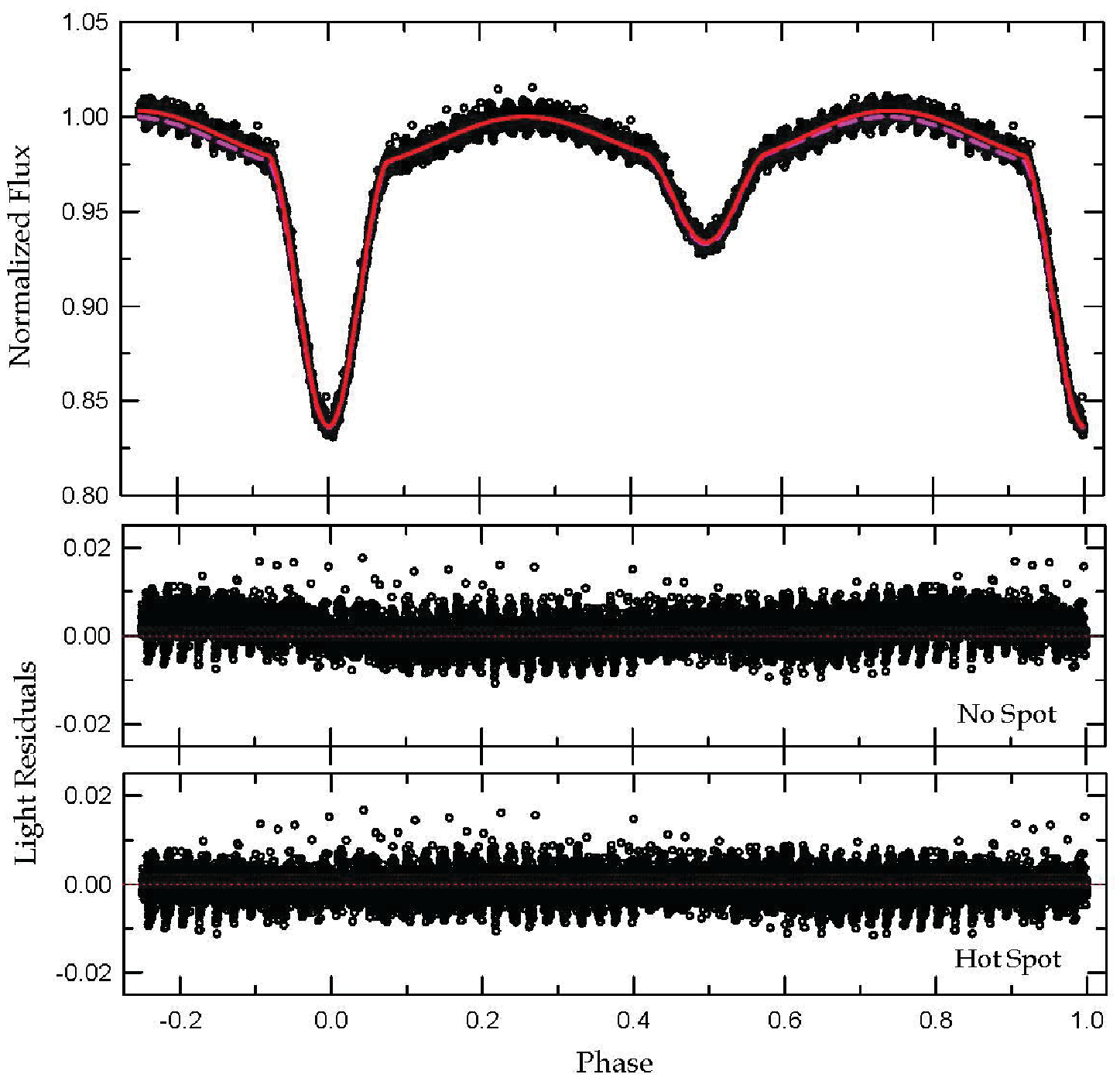}
\caption{Light curve of KIC 4739791 with the fitted models. The circles are individual measures from the {\it Kepler} spacecraft, 
and the dashed and solid curves represent the synthetic curves obtained from no spot and the hot-spot model on the primary star, 
respectively. The light residuals corresponding to the unspotted and hot-spot models are plotted in the middle and bottom panels, 
respectively. }
\label{Fig1}
\end{figure}

\begin{figure}
\includegraphics[]{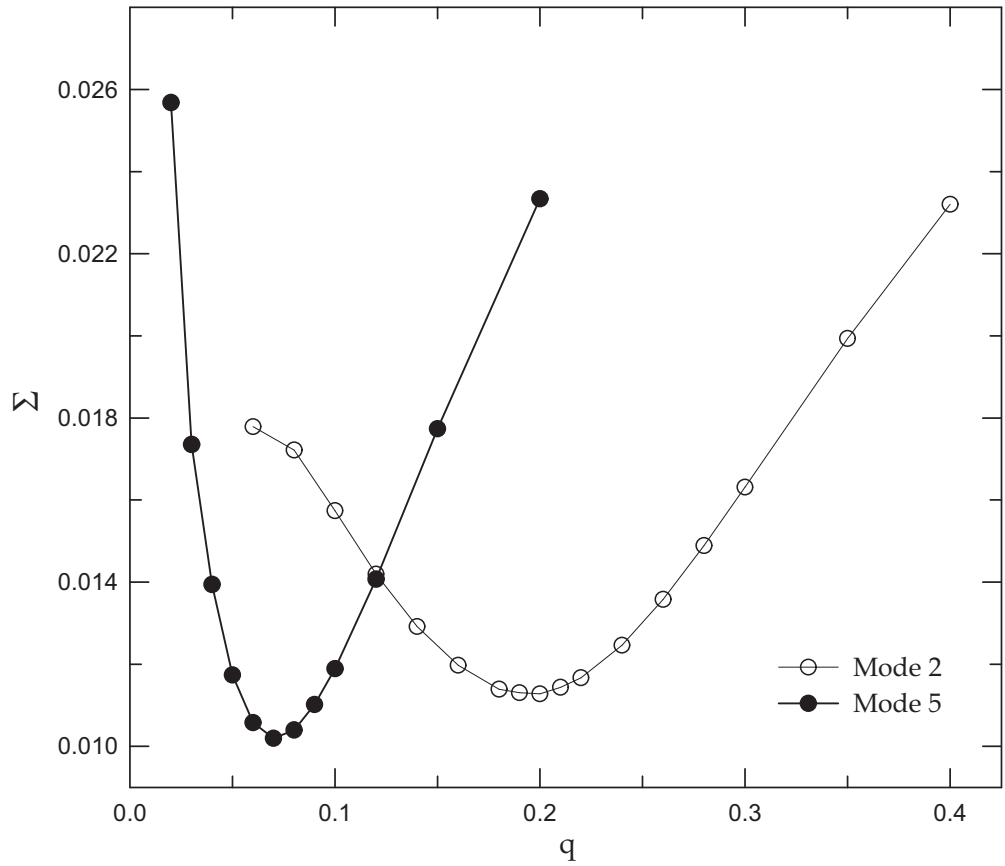}
\caption{Behavior of $\Sigma$ as a function of the mass ratio $q$. The open and filled circles represent the $q$-search results 
for the detached and semi-detached modes, respectively. }
\label{Fig2}
\end{figure}

\begin{figure}
\includegraphics[]{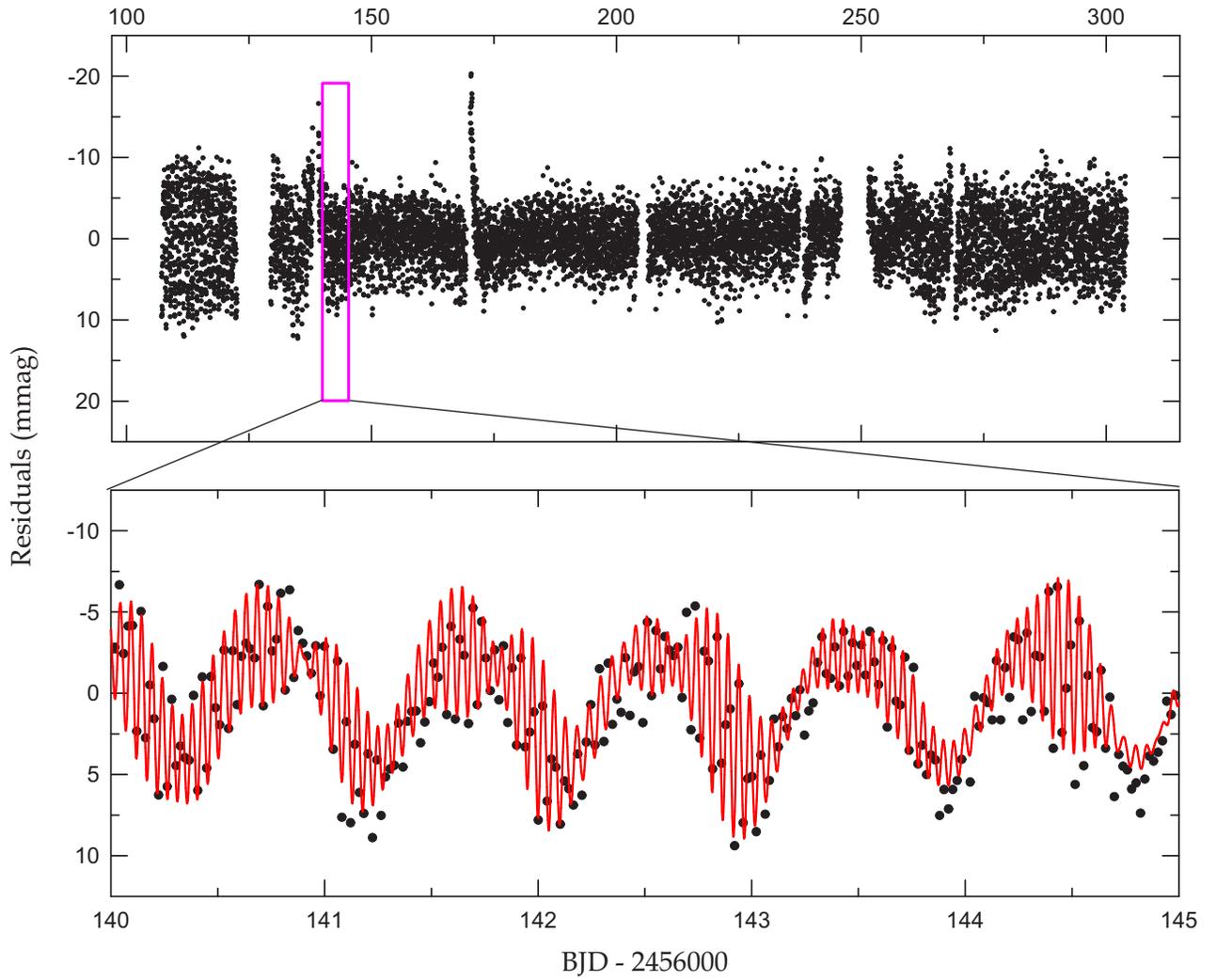}
\caption{Light curve residuals distributed in BJD. The lower panel presents a short section of the residuals marked using 
the inset box of the upper panel. The synthetic curve was computed from the 14-frequency fit to the data. }
\label{Fig3}
\end{figure}

\begin{figure}
\includegraphics[]{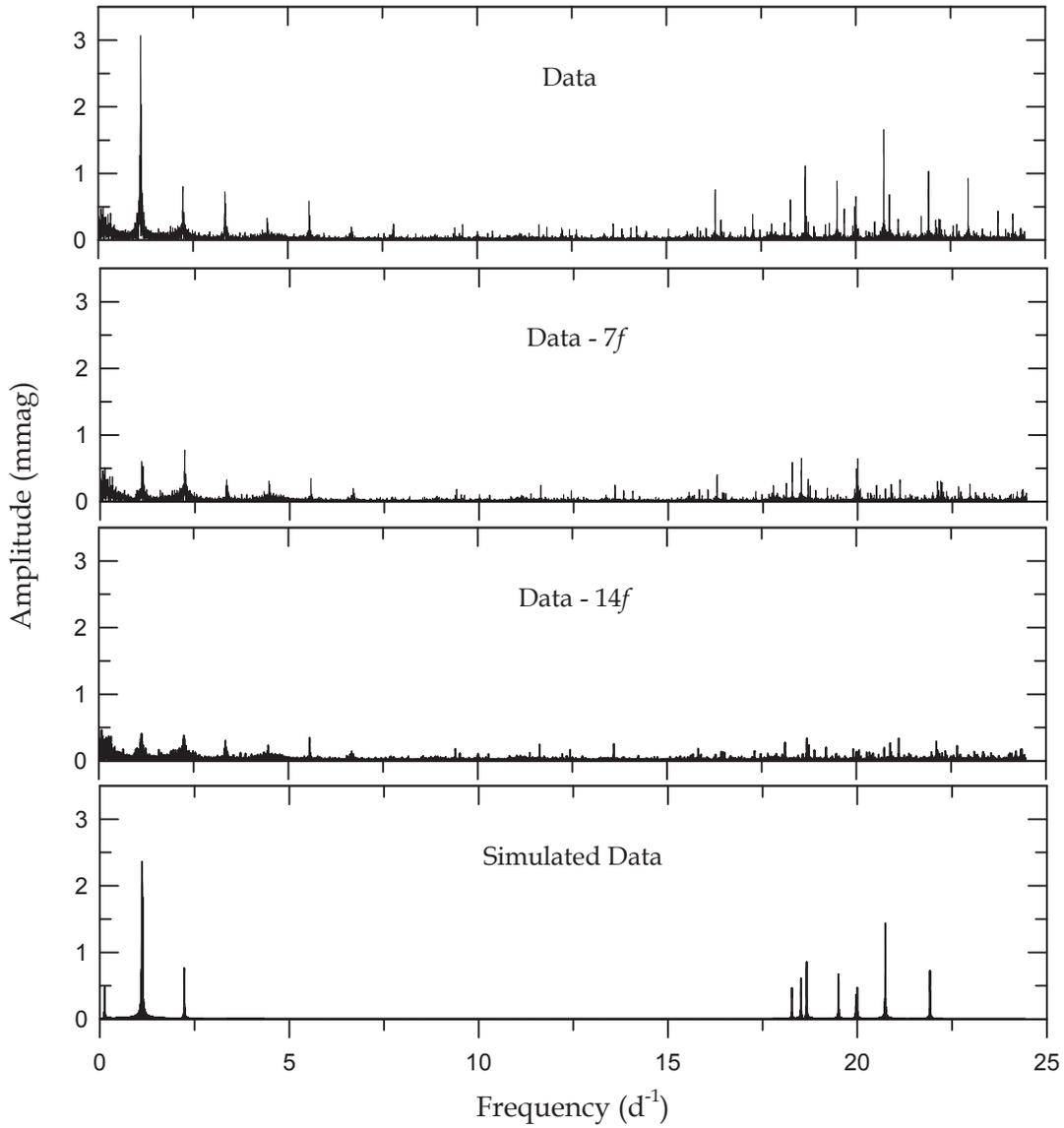}
\caption{Amplitude spectra before (top panel) and after pre-whitening the first seven frequencies (second) and all 14 
frequencies (third) from the PERIOD04 program for the outside-eclipse residual data. The periodogram for the simulated data 
is shown in the bottom panel. See the text for more detail. }
\label{Fig4}
\end{figure}

\begin{figure}
\includegraphics[]{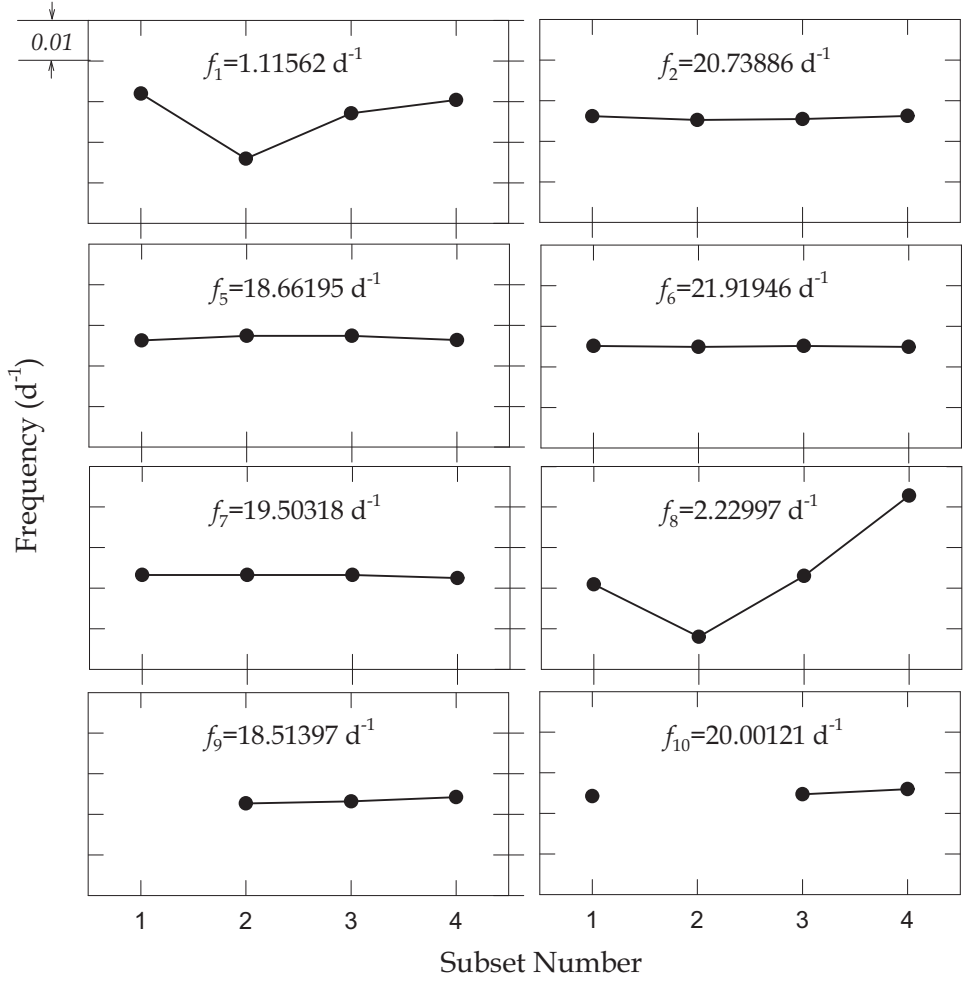}
\caption{Variability of the main frequencies detected in the four subsets at intervals of approximately 50 d. In all panels, 
the y-axes are scaled to 0.05 d$^{-1}$ and the tick intervals are 0.01 d$^{-1}$. }
\label{Fig5}
\end{figure}

\begin{figure}
\includegraphics[]{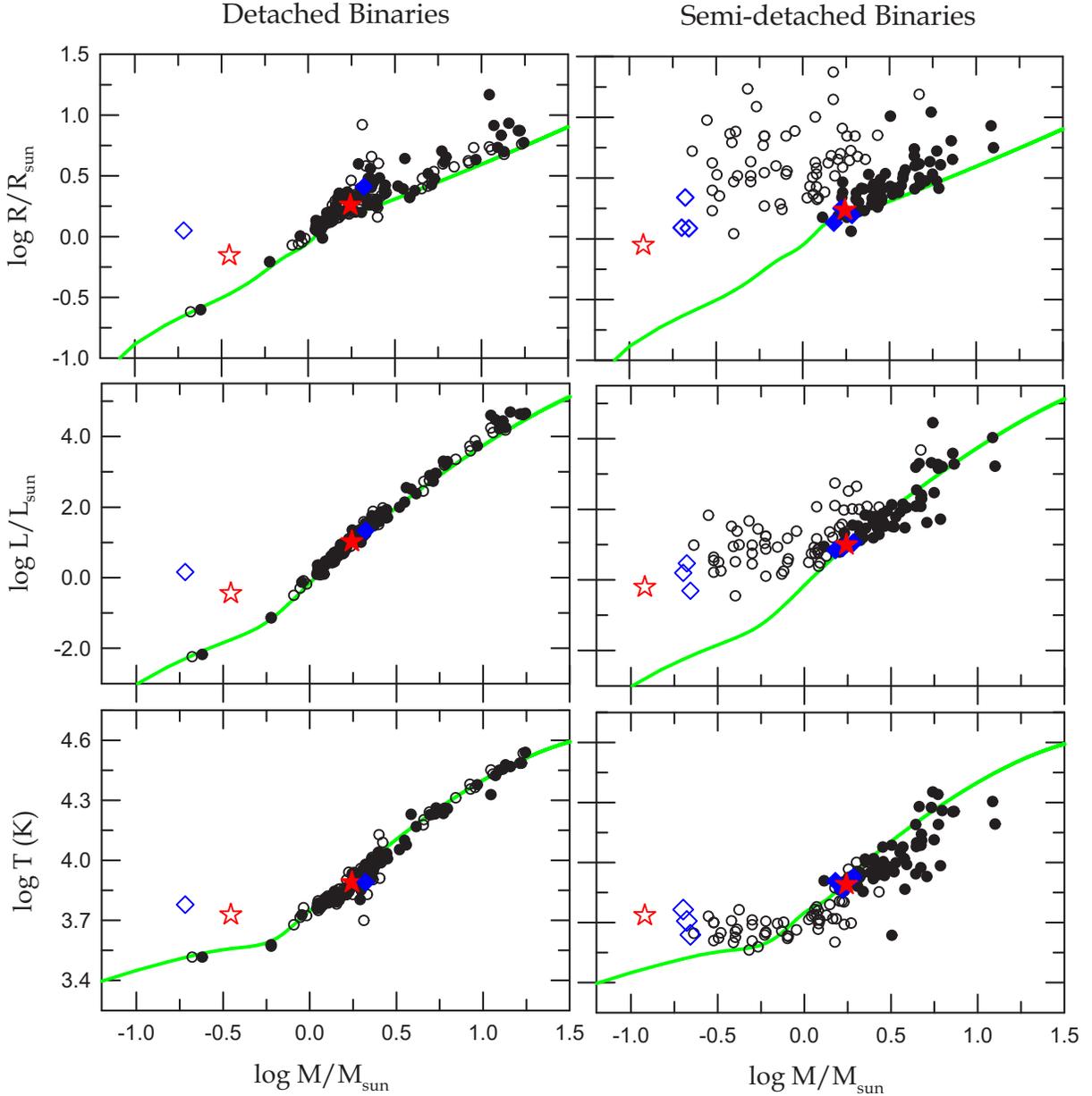}
 \caption{Mass-radius (top), mass-luminosity (middle), and mass-temperature (bottom) diagrams for the detached (left) and 
 semi-detached (right) Algols. The filled and open circles represent the primary and secondary components, respectively. 
 The star symbols denote the locations of the components of KIC 4739791 and the diamonds are the R CMa-type binaries. 
 The solid lines represent the ZAMS stars calculated as having a solar metallicity of $Z$=0.02 in Tout et al. (1996). }
\label{Fig6}
\end{figure}

\clearpage
\begin{deluxetable}{lcccccccc}
\tabletypesize{\small}  
\tablewidth{0pt} 
\tablecaption{Physical Parameters of KIC 4739791}
\tablehead{
\colhead{Parameter}                      & \multicolumn{2}{c}{Detached Mode}           && \multicolumn{5}{c}{Semi-detached Mode}                                                     \\ [1.0mm] \cline{2-3} \cline{5-9} \\[-2.0ex]
                                         & \multicolumn{2}{c}{Cool-spot Model}         && \multicolumn{2}{c}{Hot-spot Model}          && \multicolumn{2}{c}{Cool-spot Model}         \\ [1.0mm] \cline{2-3} \cline{5-6} \cline{8-9} \\[-2.0ex]
                                         & \colhead{Primary} & \colhead{Secondary}     && \colhead{Primary} & \colhead{Secondary}     && \colhead{Primary} & \colhead{Secondary}                                                  
}                                                                                                                                                                                    
\startdata                                                                                                                                                                           
$T_0$ (BJD)                              & \multicolumn{2}{c}{2,456,203.54381(6)}      && \multicolumn{2}{c}{2,456,203.54362(6) }     && \multicolumn{2}{c}{2,456,203.54403(6)}      \\
$P$ (day)                                & \multicolumn{2}{c}{0.8989195(9)}            && \multicolumn{2}{c}{0.8989195(9)}            && \multicolumn{2}{c}{0.8989199(9)}            \\
$q$                                      & \multicolumn{2}{c}{0.1995(2)}               && \multicolumn{2}{c}{0.0699(2)}               && \multicolumn{2}{c}{0.0697(2)}               \\
$i$ (deg)                                & \multicolumn{2}{c}{76.19(2)}                && \multicolumn{2}{c}{72.56(2)}                && \multicolumn{2}{c}{72.58(2)}                \\
$T$ (K)                                  & 7,751(46)         & 5,350(22)               && 7,778(28)         & 5,447(17)               && 7,719(59)         & 5,493(29)               \\
$\Omega$                                 & 3.010(2)          & 2.845(1)                && 2.900(3)          & 1.861                   && 2.873(3)          & 1.860                   \\
$\Omega_{\rm in}$                        & \multicolumn{2}{c}{2.232}                   && \multicolumn{2}{c}{1.861}                   && \multicolumn{2}{c}{1.860}                   \\
$A$                                      & 1.0               & 0.5                     && 1.0               & 0.5                     && 1.0               & 0.5                     \\
$g$                                      & 1.0               & 0.32                    && 1.0               & 0.32                    && 1.0               & 0.32                    \\
$X$, $Y$                                 & 0.673, 0.203      & 0.645, 0.183            && 0.673, 0.204      & 0.643, 0.196            && 0.673, 0.204      & 0.643, 0.196            \\
$x$, $y$                                 & 0.620, 0.251      & 0.718, 0.212            && 0.619, 0.252      & 0.707, 0.234            && 0.619, 0.252      & 0.707, 0.234            \\
$L$/($L_{1}$+$L_{2}$)                    & 0.9641(4)         & 0.0359                  && 0.9322(4)         & 0.0678                  && 0.9297(3)         & 0.0703                  \\
$r$ (pole)                               & 0.3544(3)         & 0.1372(1)               && 0.3529(4)         & 0.1703(1)               && 0.3562(4)         & 0.1702(1)               \\
$r$ (point)                              & 0.3727(3)         & 0.1416(1)               && 0.3644(4)         & 0.2550(2)               && 0.3682(4)         & 0.2548(2)               \\
$r$ (side)                               & 0.3646(3)         & 0.1383(1)               && 0.3618(4)         & 0.1771(1)               && 0.3654(4)         & 0.1769(1)               \\
$r$ (back)                               & 0.3692(3)         & 0.1410(1)               && 0.3633(4)         & 0.2069(1)               && 0.3670(4)         & 0.2067(1)               \\
$r$ (volume)                             & 0.3628(3)         & 0.1389(1)               && 0.3593(4)         & 0.1847(1)               && 0.3629(4)         & 0.1845(1)               \\ [1.0mm]
\multicolumn{6}{l}{Spot parameters:}                                                                                                                                                 \\        
Colatitude (deg)                         & \dots             & 70.0(6)                 && 88.6(8)           & \dots                   && \dots             & 75(1)                   \\        
Longitude (deg)                          & \dots             & 103(2)                  && 66(1)             & \dots                   && \dots             & 93(2)                   \\        
Radius (deg)                             & \dots             & 20(1)                   && 12.6(6)           & \dots                   && \dots             & 20.3(7)                 \\        
$T$$\rm _{spot}$/$T$$\rm _{local}$       & \dots             & 0.80(2)                 && 1.024(2)          & \dots                   && \dots             & 0.905(6)                \\
$\Sigma W(O-C)^2$                        & \multicolumn{2}{c}{0.00342}                 && \multicolumn{2}{c}{0.00327}                 && \multicolumn{2}{c}{0.00335}                 \\ [1.0mm]
\multicolumn{6}{l}{Absolute parameters:}                                                                                                                                             \\            
$M$ ($M_\odot$)                          & 1.75(8)           &  0.35(2)                && 1.75(8)           &  0.122(6)               && 1.75(8)           &  0.122(6)               \\
$R$ ($R_\odot$)                          & 1.82(3)           &  0.70(1)                && 1.73(3)           &  0.89(2)                && 1.75(3)           &  0.89(2)                \\
$\log$ $g$ (cgs)                         & 4.16(2)           &  4.30(3)                && 4.20(3)           &  3.63(3)                && 4.19(3)           &  3.63(3)                \\
$L$ ($L_\odot$)                          & 11(1)             &  0.36(5)                && 10(1)             &  0.63(9)                && 10(1)             &  0.65(9)                \\
$M_{\rm bol}$ (mag)                      & 2.2(1)            &  5.9(2)                 && 2.3(1)            &  5.2(2)                 && 2.3(1)            &  5.2(2)                 \\
\enddata
\tablenotetext{a}{Mean volume radius.}
\end{deluxetable}

\begin{deluxetable}{lrcccccc}
\tablewidth{0pt}
\tablecaption{Multiple Frequency Analysis of KIC 4739791$\rm ^a$}
\tablehead{
             & \colhead{Frequency}    & \colhead{Amplitude} & \colhead{Phase} & \colhead{S/N$\rm ^b$}  & \colhead{$Q_D$}  & \colhead{$Q_{SD}$} & \colhead{Remark}           \\
             & \colhead{(day$^{-1}$)} & \colhead{(mmag)}    & \colhead{(rad)} &                        & \colhead{(days)} & \colhead{(days)}   &
}                                                                                                                                            
\startdata                                                                                                                                   
$f_{1}$      &  1.11562$\pm$0.00009   & 2.22$\pm$0.09       & 1.68$\pm$0.04   & 18.49                  &                  &                    & $f_{\rm orb}$(?)           \\
$f_{2}$      & 20.73886$\pm$0.00005   & 1.97$\pm$0.05       & 4.76$\pm$0.02   & 35.24                  & 0.026            & 0.028              &                            \\
$f_{3}$      &  1.12095$\pm$0.00015   & 1.35$\pm$0.15       & 4.10$\pm$0.07   & 11.21                  &                  &                    & $f_1+$0.00533              \\
$f_{4}$      &  1.11029$\pm$0.00013   & 1.61$\pm$0.13       & 5.19$\pm$0.06   & 13.41                  &                  &                    & $f_1-$0.00533              \\
$f_{5}$      & 18.66195$\pm$0.00008   & 1.10$\pm$0.08       & 5.04$\pm$0.04   & 21.34                  & 0.029            & 0.031              &                            \\
$f_{6}$      & 21.91946$\pm$0.00009   & 1.03$\pm$0.09       & 2.30$\pm$0.04   & 19.46                  & 0.025            & 0.027              &                            \\
$f_{7}$      & 19.50318$\pm$0.00011   & 0.88$\pm$0.10       & 5.84$\pm$0.05   & 16.44                  & 0.028            & 0.030              &                            \\
$f_{8}$      &  2.22997$\pm$0.00021   & 0.77$\pm$0.21       & 3.24$\pm$0.10   &  8.07                  &                  &                    & $2f_1$                     \\
$f_{9}$      & 18.51397$\pm$0.00011   & 0.78$\pm$0.11       & 5.27$\pm$0.05   & 15.30                  & 0.029            & 0.031              &                            \\
$f_{10}$     & 20.00121$\pm$0.00015   & 0.62$\pm$0.15       & 1.78$\pm$0.07   & 11.38                  & 0.027            & 0.029              &                            \\
$f_{11}$     &  1.09633$\pm$0.00035   & 0.60$\pm$0.34       & 5.91$\pm$0.16   &  4.97                  &                  &                    &                            \\
$f_{12}$     & 18.27333$\pm$0.00014   & 0.60$\pm$0.14       & 2.66$\pm$0.06   & 12.44                  & 0.029            & 0.032              & $2f_{10} - f_7 - f_8$      \\
$f_{13}$     &  0.12489$\pm$0.00043   & 0.50$\pm$0.42       & 0.75$\pm$0.20   &  4.04                  &                  &                    & $f_2 - f_7 - f_1$          \\
$f_{14}$     & 19.96415$\pm$0.00020   & 0.47$\pm$0.20       & 5.62$\pm$0.09   &  8.66                  & 0.027            & 0.029              & $f_{10} + 2f_{11} - f_8$   \\
\enddata                                                                                                                           
\tablenotetext{a}{Frequencies are listed in order of detection. }
\tablenotetext{b}{Calculated in a range of 5 d$^{-1}$ around each frequency. }
\end{deluxetable}

\end{document}